\documentclass[sigplan,nonacm]{acmart}
\setcitestyle{numbers,sort&compress}
\renewcommand\footnotetextcopyrightpermission[1]{}
\copyrightyear{2025}
\acmYear{2025}
\setcopyright{acmcopyright}\acmConference[ASPLOS]{ACM International Conference on Architectural Support for Programming Languages and Operating Systems}{Pittsburgh,USA}

\usepackage{graphicx}
\usepackage{xcolor}
\usepackage{algorithm}
\usepackage{algpseudocode}
\usepackage{extarrows}
\usepackage{stackengine}
\usepackage{tikz}
\usepackage{listings}
\usepackage{graphicx}
\usepackage{textcomp}
\usepackage{soul}
\usepackage{wrapfig}
\usepackage[font=footnotesize,labelfont=bf]{caption}
\usepackage{multirow}
\usepackage{algpseudocode}
\usepackage{algorithm}
\usepackage{colortbl}

\usepackage{pifont}
\newcommand{\Circle}[1]{\raisebox{.5pt}{\textcircled{\raisebox{-.9pt}{\textbf{\fontsize{7pt}{6pt}\selectfont #1}}}}}

\usepackage{tikz}

\usepackage{tikz}
\author{Arsalan Ali Malik}
\email{aamalik3@ncsu.edu}
\affiliation{%
  \institution{North Carolina State University}
  \state{North Carolina}
  \country{USA}
}
\author{John Buchanan}
\email{jlbucha4@ncsu.edu}
\affiliation{%
  \institution{North Carolina State University}
  \state{North Carolina}
  \country{USA}
}

\author{Aydin Aysu}
\email{aaysu@ncsu.edu}
\affiliation{%
  \institution{North Carolina State University}
  \state{North Carolina}
  \country{USA}
}

\begin{document}
\title{Preemption-Enhanced Benchmark Suite for FPGAs}

\begin{abstract} 
Field-Programmable Gate Arrays (FPGAs) have become essential in cloud computing due to their reconfigurability, energy efficiency, and ability to accelerate domain-specific workloads. As FPGA adoption grows, research into task scheduling and preemption techniques has intensified. However, the field lacks a standardized benchmarking framework for consistent and reproducible evaluation. Many existing studies propose innovative scheduling or preemption mechanisms but often rely on proprietary or synthetic benchmarks, limiting generalizability and making comparison difficult. This methodical  fragmentation hinders effective evaluation of scheduling strategies and preemption in multi-tenant FPGA environments.

This paper presents the \textit{first} open-source preemption-enabled benchmark suite for evaluating FPGA preemption strategies and testing new scheduling algorithms, \emph{without} requiring users to create preemption workloads from scratch. The suite includes $27$ diverse applications spanning cryptography, AI/ML, computation-intensive workloads, communication systems, and multimedia processing. Each benchmark integrates comprehensive context-saving and restoration mechanisms, facilitating reproducible research and consistent comparisons. Our suite not only simplifies testing FPGA scheduling policies but also benefits OS research by enabling the evaluation of scheduling fairness, resource allocation efficiency, and context-switching performance in multi-tenant FPGA systems, ultimately supporting the development of better operating systems and scheduling policies for FPGA-based environments. We also provide guidelines for adding new benchmarks, enabling future research to expand and refine FPGA preemption and scheduling evaluation.
\end{abstract}

\maketitle

\section{Introduction}\label{Sec:Introduction}
Field-programmable gate arrays (FPGAs) are widely recognized for their versatility and performance in various computational domains, primarily due to their reconfigurability and parallel processing capabilities. These features have also made FPGAs an attractive solution for integration into cloud environments, enhancing productivity while reducing infrastructure costs for end-users concurrently~\cite{leeser2021fpgas}. Despite their growing cloud adoption, FPGAs lack native, vendor-supported mechanisms for pausing running designs, capturing their state, and resuming execution---known as preemption. Preemption has been shown to improve resource utilization by enabling efficient workload balancing across cloud-based compute resources such as CPUs and GPUs~\cite{Coyote}. Despite this proven value in traditional cloud environments, preemption remains unsupported in FPGA-based systems by both vendors and cloud service providers.
\\\indent
To address this gap, researchers over the last decade have proposed several custom solutions allowing FPGA designs to pause, save, and resume on demand, as shown in Fig.~\ref{fig:Trend}. This rising trend demonstrates that researchers are increasingly focusing on enabling preemption in FPGAs~\cite{StateMover,malik2025epoch, StopnLook, StateMover2, Coyote, ATC_compiler_1, ATC_hwctxsw_2, AmorphOS, OPTIMUS, STFS, Nimblock}. This growing attention also highlights the need for dynamic and flexible execution models in FPGA-based systems. However, two challenges exist with these solutions: 
\begin{figure}[t!]
    \centering
    \vspace{1.20em}
     \includegraphics[width =\columnwidth]{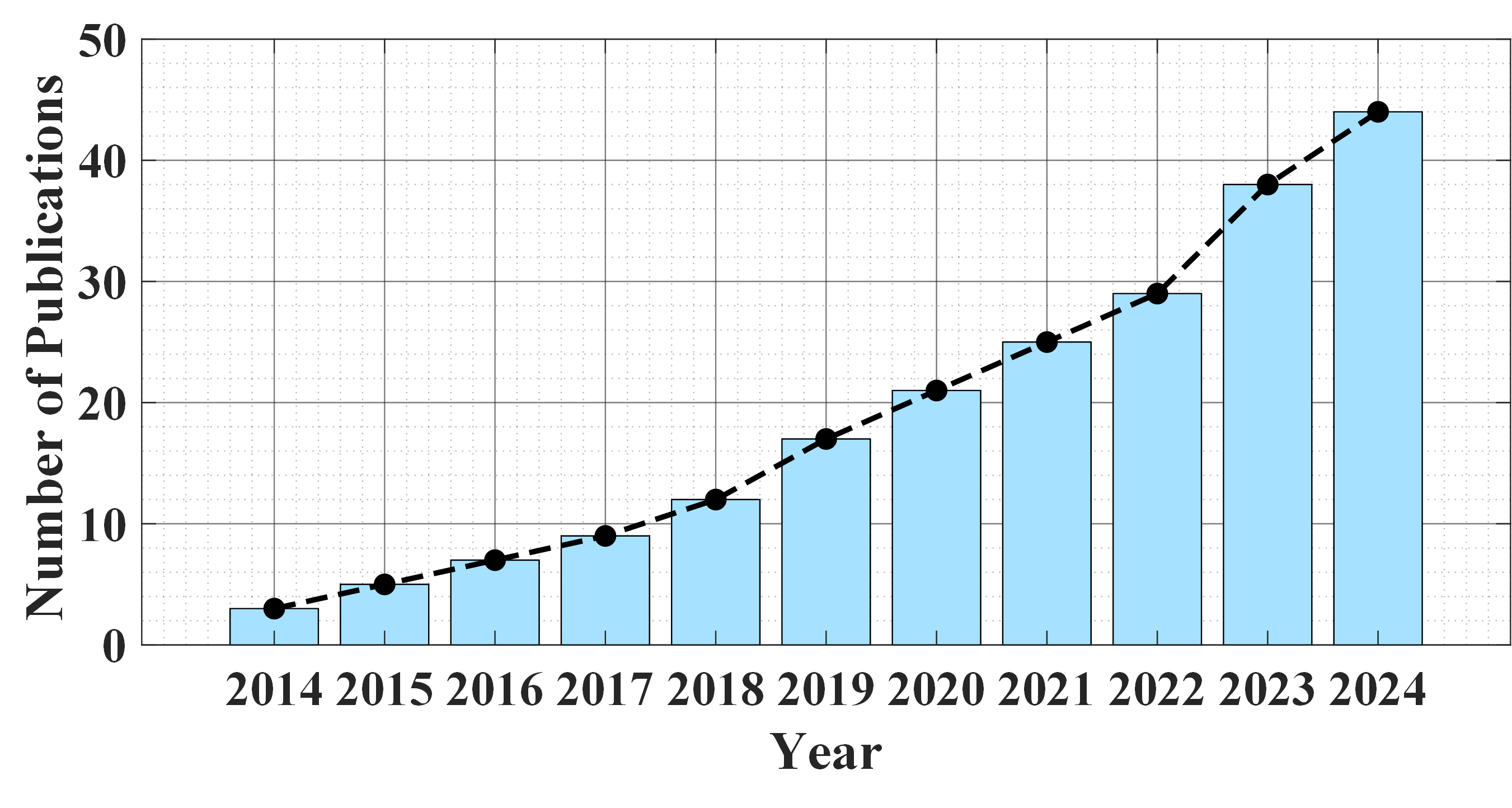}
     \vspace{-2.0em}
        \caption{Temporal distribution of academic publications referencing `FPGA preemption' and `FPGA context-switching scheduling' over the past decade, as indexed by Google Scholar~\cite{StateMover,malik2025epoch, StopnLook, StateMover2, Coyote, ATC_compiler_1, ATC_hwctxsw_2, AmorphOS, OPTIMUS, STFS, Nimblock}. The data exhibits a clear increase—reflecting the industrial and academic interest—in ongoing research efforts in managing FPGA resources dynamically through the preemption capabilities.}
        \label{fig:Trend}
\vspace{-1.75em}
\end{figure}
(i) The existing solutions have primarily been tested using small-scale, toy examples, which do not scale gracefully to full systems or large workloads. (ii) Most of these solutions are not open-sourced, thereby limiting their accessibility, testing, and further development by the broader research community. Consequently, there is a clear need for a preemption-enabled FPGA benchmark suite that is open-source, scalable, and representative of a diverse range of applications across fields, such as cryptography, artificial intelligence (AI), machine learning (ML), communication systems, image, video and signal processing.
\\\indent
Benchmarking is indeed critical in embedded systems research, as it enables reproducible evaluation of performance, resource utilization, and system behavior under diverse workloads. Moreover, it allows for the fair and consistent comparison of multiple competing approaches, guiding researchers in systematically identifying strengths, weaknesses, and suitability tailored to specific use cases. Established benchmark suites, such as MiBench for embedded processors~\cite{Guthaus2001mibench}, CoreMark for microcontrollers~\cite{gal2012exploring}, and domain-specific abstractions like the Seven and Twelve Dwarfs~\cite{krommydas2016opendwarfs} have guided architectural and software development across real-time systems, sensor networks, and edge computing. These efforts highlight the necessity of representative, standardized workloads for validating both functional and non-functional system properties.
\\\indent
Despite a proliferation of methodologies for FPGA task scheduling and preemption, a \emph{unified} benchmarking support remains fragmented and inconsistent. Existing evaluations—such as STFS, Coyote, ReconOS, and others—often rely on ad hoc or narrowly scoped workloads, which impedes cross-method comparison and reproducibility~\cite{STFS, Coyote, Virtex6_preemptive, Arch_Support, PR_OpenCL}. The benchmarks employed by these works lack standardization, breadth across application domains, and consistent metrics. For instance, STFS uses synthetic workloads with limited diversity, Coyote demonstrates capabilities on a few selected applications without integration into a standardized suite, and ReconOS focuses on feasibility rather than complexity. \\\indent
Consequently, it remains unclear (i) which scheduling method offers superior performance and (ii) whether reported research outcomes \emph{accurately} represent real-world FPGA usage and scenarios. To address these gaps, we introduce a suite of preemptible workloads specifically designed for FPGA-based systems. Our standardized benchmark suite enables rigorous evaluation of interruptibility, context-switch overhead, and runtime integration, allowing for fair comparison and ranking of competing scheduling approaches. By supporting both offline and online preemption across a diverse set of $27$ real-world benchmarks, our suite establishes the first comprehensive, preemption-aware framework for FPGAs, ensuring that benchmarking results closely reflect practical deployment scenarios.\footnote{The choice of benchmarks employed in this work reflects real-world workloads, similar to the OpenDwarfs benchmark suite used in general-purpose computing~\cite{krommydas2016opendwarfs}.}

The primary contribution of this paper is the introduction of the \textit{first} open-source preemption-capable benchmark suite FPGAs that:
\begin{itemize}
    \item Includes a diverse set of benchmarks representing real-world applications in fields such as cryptography, artificial intelligence/machine learning (AI/ML), computational workloads, communication systems, video, image, and signal processing.
    
    \item Enhances cloud FPGA capabilities by integrating previously proven and real-world tested work. 

    \item Provides an open-source, scalable solution that can be used by the broader research community to evaluate preemption techniques on a variety of real-world workloads.
\end{itemize}
By releasing this work publicly, we aim to provide a \emph{unified} platform for comparing preemption strategies, fostering reproducibility in reconfigurable systems research.
\vspace{0.75em}
\vspace{-1.5em}
\section{The Need for a New Benchmark Suite}\label{Sec:motivation}
\vspace{-0.25em}
Effective preemption mechanisms significantly enhance resource utilization and responsiveness in cloud environments, enabling the efficient sharing of resources and improved task scheduling across heterogeneous workloads~\cite{leeser2021fpgas, Coyote, karabulut2024themis, STFS}. Although various research efforts have attempted to address FPGA preemption, a noticeable gap remains in comprehensive and realistic benchmarking frameworks~\cite{malik2025epoch, StateMover, Nimblock,StateMover2}. These works either focus solely on processor-based (soft-core CPU) scenarios or utilize overly simplistic hardware accelerators that inadequately represent the diversity and complexity of real-world applications. 

Recognizing these limitations, we introduce an open-source benchmark suite designed to fill the gap in evaluating preemption-capable FPGA workloads. This work comprises two complementary benchmark groups: programmable logic (PL)-based hardware accelerators and RISC–V processor-based accelerators\footnote{In this work, the terms `accelerator, kernel, workload and benchmark' are used interchangeably.} that span across multiple application domains, including cryptography, artificial intelligence, signal processing, image and video processing, network-on-chip (NoC) interconnects, and general-purpose soft processors. The aim of this work is to facilitate rigorous, and reproducible research on FPGA preemption techniques by providing representative and realistic evaluation scenarios, thus enabling advances in FPGA multitasking and cloud integration.
\vspace{-1.25em}
\section{Background and Related Works}\label{Sec:background}
\vspace{-0.25em}
System-on-chip (SoC) architectures typically integrate FPGAs with two main components: the programmable logic (PL) and the processing system (PS). The PL forms the reconfigurable fabric of the FPGA and consists of an array of configurable logic blocks, interconnects, and embedded resources that can be programmed to implement custom hardware accelerators or digital circuits. Within the PL, look-up tables (LUTs) can be used to implement combinational logic functions, flip-flops (FFs) to provide storage for sequential logic, digital signal processing blocks (DSPs) to accelerate arithmetic operations such as multiplication and addition, and block RAMs (BRAMs) as on-chip memory to store data and instructions. Conversely, the PS typically contains fixed-function processors (\textit{e.g.,} ARM cores), memory controllers, and peripheral interfaces that handle control, communication, and general-purpose computation.
\begin{table*}[t]
\footnotesize
\centering
\caption{The table illustrates the range of benchmarks employed in prior works on FPGA preemption and scheduling algorithms. While some overlap exists in the selection of benchmarks, their use remains largely ad hoc, lacking systematic coverage across diverse application domains. Moreover, the limited availability of these benchmarks—many of which are either not open-source or only partially accessible—further restricts their utility and reproducibility within the broader research community.}
\vspace{-1em}
\label{tab:Prior_Benchmarks}
\begin{tabular}{|c|c|c|c|c|}
\hline
\textbf{\begin{tabular}[c]{@{}l@{}}Reference\\Works\end{tabular} }   & 
\textbf{\begin{tabular}[c]{@{}l@{}}Open-\\Source\end{tabular} }
&
\textbf{\begin{tabular}[c]{@{}l@{}}Preemption\\Capable\end{tabular} }   &
\textbf{\begin{tabular}[c]{@{}l@{}}Evaluation\\Platform \end{tabular} }   &
\textbf{\begin{tabular}[c]{@{}l@{}}Benchmarks Employed\end{tabular} }                                                                                                                                                                                                                                                                                                                                                                                                            \\ \hline
STFS~\cite{STFS}                                                          & No       & No    &  XC7Z020        & AES, BFS, SHA, SPMV, GSM, FFT, SORT, and Viterbi decoder.                                                                                                                                                                                                                                                                                                                                                                               \\ \hline
Miliadis et al~\cite{Arch_Support}                                             & No    & No   & Alveo U250            & \begin{tabular}[c]{@{}l@{}}Hotspot, K-means, KNN, Backprop(f), Pathfinder, \\Leukocyte(d), Lud, NW, FD, and Lavamd.\end{tabular}                                                                                                                                                                                                                                                                                                       \\ \hline
Vaishnav et al.~\cite{PR_OpenCL}                                            & No      & No    &  XCZU9EG        & CRC32, Matrix multiplication, and Euclidean distance (e-dist) for K-means.                                                                                                                                                                                                                                                                                                                                                              \\ \hline
Trong-Yen et al.~\cite{ATC_hwctxsw_2} & No     & No &  XC2VP50            & Up-Counter, 16/32-bit Divider, LED Display control, MLC, GCD32, DES, and DCT.                                                                                                                                                                                                                                                                                                                                                           \\ \hline
Sameh et al.~\cite{StateMover}          & No    &Yes      &  \begin{tabular}[c]{@{}l@{}}XCKU040\\ \& Arria 10 \end{tabular}        & AES, Ethernet                                                                                                                                                                                                                                                                                                                                                                                                                         \\ \hline
StateMover v1.0~\cite{StateMover2}      & No     &Yes        &  XCKU040        & Crossbar, FIR Filter, Network Sorter, AES, Crossbar-big                                                                                                                                                                                                                                                                                                                                                                                 \\ \hline
EPOCH~\cite{malik2025epoch}                                                         & No  & Yes&  XC7Z020               & \begin{tabular}[c]{@{}l@{}}AES-128, SHA-256, FALCON (Keygen, Signature Generation \& Verification),\\ GEMM, BFS, NW, KMP, SORT, Dhrystone and RISC-V Coremark.\end{tabular}                                                                                                                                                                                                                                                            \\ \hline

StateMover v2.0~\cite{StopnLook}        & Partial  & Yes   &  XCKU040        & Counter, Memcached, Digital Up-Converter, DFT-64/1024, and 64-Floating-point                                                                                                                                                                                                                                                                                                                                                            \\ \hline
ReconOS~\cite{Virtex6_preemptive}                                                       & Partial      & Yes & XC6VLX240T       & Addition, Subtraction, Multiplication, and LFSR.                                                                                                                                                                                                                                                                                                                                                                                        \\ \hline
AMORPH$_{OS}$~\cite{AmorphOS}             & Partial  & No     & Amazon F1        & \begin{tabular}[c]{@{}l@{}}Convolutional neural network,  Memory streaming, Bitcoin hashing accelerator, \\AES, DES, SHA, Double-precision (addition, multiplication, and Sine function), MIPS \\processor, Adaptive differential pulse codec, Linear predictive coding analysis,\\ JPEG image decompression, and Motion vector decoding.\end{tabular} \\ \hline

Coyote~\cite{Coyote}                                                        & Yes   &No     & XCVU9P         & \begin{tabular}[c]{@{}l@{}}HyperLogLog estimation, AES encryption, SHA256 hash, and K-means calculation.\end{tabular}                                                                                                                                                                                                                                                                                                     \\ \hline
SYNERGY~\cite{ATC_compiler_1}         & Yes       & No   &  DE10 Nano       & \begin{tabular}[c]{@{}l@{}}Pulse-code modulation encoder/decoder, Bitcoin mining, Double-precision arithmetic \\circuits, Bubble-sort on a 32-bit MIPS processor, DNA sequence alignment (NW),\\ and Streaming regular expression matcher (regex).\end{tabular}                                                                                                                                                                       \\ \hline
OPTIMUS~\cite{OPTIMUS}                  & Yes   & No    &  Arria 10          & \begin{tabular}[c]{@{}l@{}}AES-128, SHA-512, MD-5, FIR Filter, Gaussian RNG, Reed Solomon Decoder, Image\\ filters (Gaussian, Grayscale, and Sobel), Smith-Waterman algorithm, Bitcoin Miner, \\Single source shortest path, Random memory access, and linked-list walker.\end{tabular}                                                                                                                                                 \\ \hline
\textbf{This Work}~     & \textbf{Yes}      & \textbf{Yes}   &  XC7Z020         & \textbf{\begin{tabular}[c]{@{}l@{}}Neural network convolution layer, FFT, FIR Filter, IIR Filter, H.264 Video \\Encoder, Image-Processor, JPEG Decoder, PNG Decoder, MIPS-processor, \\Matrix-multiplication, ML-KEM Client \& Server, Open Network-on-Chip, \\Trigonometry core, Viterbi Decoder, AES-128, SHA-256, Merge Sort,\\ FALCON (Key generation, signature generation \& verification), GEMM, BFS, \\NW, KMP, Dhrystone and RISC-V Coremark.\end{tabular} }                 \\ \hline
\end{tabular}
\vspace{-1.5em}
\end{table*}
In the FPGA context, preemption refers to the ability to temporarily pause the execution of a hardware design instantiated in the PL, save its current state—including registers, memory contents, and computation progress—and later resume execution from that exact point. This process requires reading the state of each FF, LUT, DSP block, and other relevant elements. Designers typically achieve this by accessing the FPGA configuration memory frame-by-frame. A frame in an FPGA refers to the smallest addressable unit of the device’s configuration memory (see Section~\ref{Overhead} for more details). By reading or writing configuration frames, one can capture or modify the precise state of the FPGA’s logic fabric at a fine granularity. Capturing and restoring these frames enables precise preservation and recovery of the entire design state. 
\\\indent
Preemption in FPGA systems has received growing attention in recent years, yet existing efforts vary significantly in their practicality, scope, and scalability~\cite{STFS,Coyote,karabulut2024themis,Virtex6_preemptive,Arch_Support,PR_OpenCL}. Despite numerous methodologies proposed for FPGA-based task scheduling and preemption, existing benchmarking support remains insufficient. Notably, there is a lack of \emph{unified},  comprehensive benchmarks to facilitate fair and consistent comparisons of multiple, competing scheduling approaches \textit{e.g.,} STFS employs non-preemptive scheduling with periodic reallocation of FPGA slots~\cite{STFS}. However, STFS's evaluation is limited to a narrow set of synthetic workloads, lacking diversity in application domains and complexity levels. Coyote introduces a three-layer hierarchical design supporting dynamic partial reconfiguration~\cite{Coyote}; however, its evaluation relies on task-specific benchmarks that are not part of a standardized suite. This narrow, task-specific evaluation prevents meaningful comparison with other methodologies and restricts the ability to assess the generality or effectiveness of their scheduling approach across diverse application domains.
\\\indent
ReconOS utilizes the Virtex-6 FPGA for preemptive scheduling, focusing on demonstrating the feasibility of context switching~\cite{Virtex6_preemptive}. However, ReconOS employs minimalistic designs, such as simple counters or state machines, which do not capture the complexity and intricacies of real-world applications, thereby limiting ReconOS's applicability. Similarly, in another work,  architectural support for preemption in FPGAs has been explored, but their evaluations frequently rely on custom, non-standard benchmarks~\cite{Arch_Support}. Other efforts to enable preemption in high-level synthesis environments such as OpenCL also tend to rely on ad hoc workloads~\cite{PR_OpenCL}. 
\\\indent
The lack of standardized evaluation metrics and benchmark classes across these studies fundamentally impedes rigorous, consistent assessment of FPGA scheduling techniques\footnote{We refer the motivated readers to the following surveys for further details on the trends in `benchmarking' and `scheduling'~\cite {njuguna2008survey,reuther2019survey,kuon2008fpgasurvey}.}. 
Table~\ref{tab:Prior_Benchmarks} illustrates the variety of benchmarks employed in prior studies, revealing that while there is some overlap in benchmark selection, their use has been largely ad hoc and lacks systematic coverage across different application domains. While the works in~\cite{StateMover,StateMover2,malik2025epoch,StopnLook,Virtex6_preemptive} claim practical preemption support, these benchmarks are not (fully) open-sourced. The remaining studies either provide only theoretical support for preemption or offer benchmarks that are not open-source or only partially available, which hampers reproducibility, scalability, and constrains meaningful comparative evaluation within the FPGA preemption research community. By contrast, our work implements practical preemption support and releases the entire benchmark suite as open-source, ensuring accessibility for reproducibility and comparative studies.
We now provide a broad categorization of prior works according to the methodologies employed and abstraction layers targeted.
\vspace{-0.5em}
\subsection{Checkpointing and Debugging Mechanisms}
Attia and Betz explored FPGA checkpointing and context switching with an emphasis on incremental state capture for debugging~\cite{StateMover,StateMover2,StopnLook}. While pioneering, these approaches introduced significant design complexity, necessitated substantial manual intervention, and relied on the joint-task action group (JTAG) interface. Since JTAG requires physical access to the device, this dependency limits the applicability of their methods in cloud-based environments where such access is unavailable. Additionally, their evaluations were restricted to simplified test scenarios, reducing relevance to real-world FPGA workloads.
\vspace{-0.5em}
\subsection{Compiler-Driven and OS-Level Virtualization}
Landgraf et al. proposed SYNERGY, a compiler-driven framework for FPGA virtualization~\cite{ATC_compiler_1}, and Korolija et al. introduced operating system abstractions tailored to FPGA resource sharing~\cite{Coyote}. Although these works advanced resource management, they lacked dedicated mechanisms for runtime hardware preemption and systematic hardware state preservation. In another work, Khawaja et al. developed AmorphOS, emphasizing FPGA resource allocation and isolation~\cite{AmorphOS}. However, explicit hardware-level preemption support remained rudimentary, and empirical evaluations are confined to narrow workload profiles.
\vspace{-0.5em}
\subsection{Hypervisor and System-Level Resource Management}
Ma et al. introduced a hypervisor targeting shared-memory FPGA systems to enhance virtualization~\cite{OPTIMUS}. Despite its contributions to resource partitioning, the work did not address the detailed context-saving and restoration procedures essential for hardware preemption. Likewise, the works of STFS and NIMBLOCK focused on fine-grained FPGA scheduling and long-term resource management, but they omitted explicit consideration of preemptive context handling~\cite{STFS,Nimblock}.
\vspace{-1.5em}
\subsection{Partial Reconfiguration and Hardware Context-Switching}
Lee et al. presented a dynamic partial reconfiguration methodology to enable FPGA context switches~\cite{ATC_hwctxsw_2}. Although conceptually advancing hardware context migration, their proposed approach suffered from substantial overhead associated with fine-grained state capture and lacked validation against realistic, complex accelerator workloads. Malik et al. introduced the EPOCH framework, which leverages a RISC-V soft-core processor to implement preemption on FPGA platforms~\cite{malik2025epoch}. EPOCH  demonstrated the feasibility of hardware preemption---with zero fabric overhead on the PL fabric---its evaluation mainly explored soft-core benchmarks, leaving preemption behavior and associated overheads for dedicated hardware accelerators unexamined.
\vspace{-0.5em}
\subsection{Summary and Limitations of Prior Works}
Despite notable progress in the field, prior research on FPGA preemption exhibits the following limitations:
\vspace{-0.5em}
\begin{itemize}

    \item \textbf{Limited scalability.} Experimental evaluations often employ oversimplified hardware scenarios, which can misrepresent the diversity and complexity of production FPGA applications~\cite{StateMover2,Nimblock,StateMover}.

    \item \textbf{Restricted open-source availability.} Proprietary methodologies and the \emph{lack} of open-source benchmarks hinder reproducibility and limit the scope of comparative analysis~\cite{AmorphOS,STFS,Coyote}.
 
    \item \textbf{Narrow scope.} Preemption evaluations have predominantly centered on soft processors or small-scale examples, neglecting realistic, dedicated hardware accelerators~\cite{ICAPcontext,Coyote,malik2025epoch}.
\end{itemize}
The proposed benchmark suite addresses these shortcomings by providing a comprehensive, open-source set of benchmarks for preemption evaluation. Spanning diverse domains and realistic workloads, it enables reproducible comparisons of FPGA preemption strategies and supports advances in multitasking and cloud computing.
\vspace{-0.5em}
\section{The Proposed Benchmark Suite}\label{Sec:This_Work}
To comprehensively evaluate FPGA preemption mechanisms, this work incorporates two sets of benchmarks: $15$ PL-based hardware accelerator designs spanning diverse application domains and $12$ RISC-V–based benchmarks derived from prior work~\cite{malik2025epoch}. Instead of developing benchmarks from scratch\footnote{Apart from scalable matrix multiplication (see~\ref{matrix-multiplciation}), we do not claim ownership of the benchmarks. Our contribution lies in adapting these existing, validated benchmarks to introduce necessary preemption support.},  we consolidated and built upon pre-established, well-regarded benchmarks that have been extensively tested and are widely recognized within the research community~\cite{benchmark,PL_aditya2013trigonometric,PL_valar1234_mips,PL_xing2021compact,PL_agarwal2022image,PL_bcattle_hardh264,PL_owocomm0_fpgafft,PL_reddy2019opennoc,PL_wang2021fpga_png_decoder,PL_bakhshalipour_noc_verilog,PL_wang2021fpga,PL_padhi499_fpga_cnn}. 
Our contribution is to extend these validated benchmarks by incorporating preemption support, enabling systematic evaluation without the need to redesign benchmarks from scratch.
Each benchmark is detailed below, organized by domain, with emphasis on functionality, real-world relevance, and preemption-specific considerations.
\vspace{-1.25em}
\subsection{PL-based Hardware Accelerators}
\vspace{-0.25em}
Our work includes a total of $15$ PL-based accelerator benchmarks drawn from widely used designs in the research community. 
A brief description of each PL-based benchmark is provided below:
\vspace{-0.5em}
\subsubsection{AI and Machine Learning} \label{matrix-multiplciation}This benchmark implements a fully parameterizable and scalable matrix multiplication accelerator—an essential operation in scientific computing and machine learning workloads, including neural networks, \textit{e.g.,} convolution and dense layers\footnote{The multiplication operation is lightweight and modular, providing a foundation for future integration of activation functions—such as the rectified linear unit (ReLU)—which are not currently supported.}. The design scales its parallelism through hardware pipelines and arrays of multiply-accumulate units, allowing performance-area tradeoffs. Large-scale matrix multiplications produce significant intermediate states (\textit{e.g.,} accumulators, indices), making precise state capture and correct resumption crucial yet complex, highlighting the importance of preemption support.
\vspace{-1.25em}
\subsubsection{Image and Video Processing} This set consists of PNG Decoder (tested using images 10 and 14), JPEG Decoder (tested using images 01 and 07), H.$264$ video codec, and Image Processor~\cite{PL_wang2021fpga_png_decoder,PL_wang2021fpga,PL_bcattle_hardh264,PL_agarwal2022image}. These benchmarks accelerate common multimedia workloads. The PNG and JPEG decoders implement entropy decoding and image reconstruction pipelines; H.$264$ video codec encodes raw video into H.264 format, supporting real-time video streaming; the Image Processor performs operations, \textit{e.g.,} color space conversion (brightness, contrast, threshold adjustments) and filtering. These designs address challenges such as preserving the state of processing pipelines and internal buffers during preemption. 
Multimedia processing involves intricate pipelines with complex internal buffering and data dependencies. These characteristics make state preservation during preemption essential, yet challenging for smooth and error-free operation in multimedia cloud services.
\vspace{-0.5em}
\subsubsection{Cryptography} 
This benchmark implements the module lattice-based key encapsulation mechanism (ML--KEM)\footnote{Winner of the NIST post-quantum cryptography competition.}. It adopts a server-client model to reflect emerging secure communication demands~\cite{PL_xing2021compact}. The server generates key pairs, and the client executes key encapsulation and decapsulation procedures. Cryptographic workloads like ML–KEM handle sensitive states—such as secret keys and random seeds—which require secure state-saving to maintain confidentiality and correctness upon resumption.
\vspace{-0.5em}
\subsubsection{Signal Processing and Mathematical Computation} These benchmarks consist of the FFT Accelerator and the Trigonometry Core~\cite{PL_aditya2013trigonometric,PL_owocomm0_fpgafft}.  The FFT benchmark transforms time-domain signals to the frequency domain, while the Trigonometry core computes sine and cosine functions using CORDIC or polynomial approximations. Both designs highlight challenges of iterative computations: the FFT benchmark ensures preservation of stage progress and partial results, while the Trigonometry core captures pipeline registers and iteration counters. Iterative computations in these benchmarks generate intermediate results critical for accuracy. Precise preemption support is therefore essential to ensure correctness, avoid costly recomputation, and enable efficient task resumption.

\vspace{-0.5em}\subsubsection{Networking (Network-on-Chip Fabrics)} This design implements parameterizable network-on-chip (NoC) fabrics, facilitating packet-switched communication among FPGA cores. Open NoC represents a complex mesh-based fabric~\cite{PL_reddy2019opennoc}. Complex communication patterns, involving numerous in-flight packets and distributed router states, pose intricate challenges for accurately managing preemption, emphasizing the necessity for robust state capture and seamless resumption mechanisms.

\vspace{-0.5em}\subsubsection{Soft-Core Processor} This $32$-bit MIPS soft-core benchmark demonstrates preemption in general-purpose FPGA-based CPUs~\cite{PL_valar1234_mips}. This soft core is capable of executing synthetic workloads. General-purpose processors require comprehensive architectural state capture (\textit{e.g.,} program counter, registers, caches, pipelines), introducing significant complexity in achieving seamless preemption. This benchmark complements RISC-V-based benchmarks by providing architectural diversity critical for evaluating processor-centric preemption strategies comprehensively.

\vspace{-0.5em}\subsection{RISC-V Processor-based Hardware Accelerators}
In addition to the PL-based hardware accelerators, our work also incorporates a set of RISC-V soft-core benchmarks inspired by the prior work~\cite{malik2025epoch}. This set of benchmarks executes on a cv$32$e$40$x RISC-V soft processor operating at $100$ MHz and includes algorithmic kernels from MachSuite~\cite{benchmark} as well as standard CPU performance tests from the embedded microprocessor benchmark consortium (EEMBC)~\cite{gal2012exploring}. To evaluate diverse processor workloads under preemption scenarios, we categorize the RISC-V soft-core benchmarks across five key application domains. These include cryptography, pattern recognition, scientific computation, data analytics, and general-purpose system benchmarking. A brief description of these benchmark is provided below:
\vspace{-0.5em}\subsubsection{Cryptography} These benchmarks include encryption, hashing, and digital signature generation. Cryptographic operations produce sensitive internal states vital for maintaining security. Secure state preservation during preemption is critical for safely interrupting and resuming these workloads.

\begin{itemize}
    \item Advanced encryption standard (AES) with a 128-bit key, representing a widely adopted symmetric-key encryption algorithm~\cite{benchmark2}. It stresses arithmetic operations, bitwise logic, and memory access patterns, modeling secure communication scenarios.

    \item SHA-256 cryptographic hash function, generating a fixed-length digest from variable-length inputs~\cite{RISC_conte_crypto_algorithms}. This benchmark is representative of hashing workloads found in integrity verification, digital signatures, and blockchain protocols.

    \item FALCON (Post-Quantum Digital Signatures) evaluates the FALCON
    lattice-based post-quantum digital signature algorithm~\cite{RISC_FALCON}. It includes: \begin{itemize} 
    \vspace{0.5em}
    \item \textbf{Key Generation}: Computationally intensive lattice-based public-private key construction.
    \vspace{0.5em}
    \item \textbf{Signature Generation}: Authenticates data by signing messages with a private key. 
    \vspace{0.5em}
    \item \textbf{Signature Verification}: Validates signatures using the public key and original message. \end{itemize} 
\end{itemize}

\vspace{-0.5em}\subsubsection{Pattern Recognition and String Processing} This benchmark consists of Knuth–Morris–Pratt (KMP) and Needleman–Wunsch (NW) sequence alignment~\cite{benchmark}. KMP implements a linear-time substring search algorithm for pattern detection. It exercises control flow and pointer arithmetic, representing workloads in data stream analysis and intrusion detection. By contrast, NW is a dynamic programming algorithm for optimal sequence alignment, commonly used in bioinformatics. It stresses nested loop structures and non-uniform memory access.

\vspace{-0.5em}\subsubsection{Machine Learning} General matrix-matrix multiplication (GEMM) is a core linear algebra kernel~\cite{benchmark}, widely employed in machine learning (\textit{e.g.,} fully connected layers in neural networks), numerical simulation, and high-performance computing. It is a representative workload for evaluating arithmetic intensity and memory bandwidth demands in compute-intensive applications. GEMM operations have significant computational intensity and heavy data dependencies, necessitating sophisticated mechanisms for accurate state management to ensure seamless, correct task resumption after context switching.

\vspace{-0.5em}\subsubsection{Database Analytics and Sorting}  This benchmark consists of the most commonly used database manipulation operations: breadth-first search (BFS) and sorting; specifically, merge sort (MS)~\cite{benchmark}. BFS implements a graph traversal algorithm that explores nodes in a level-order (queue) fashion. This benchmark reflects graph analytics workloads, stressing irregular memory access and control logic. By contrast, MS sorts integer arrays using the merge sort algorithm. These algorithms exhibit irregular data access patterns, dynamic memory interactions, and complex control flows. Such characteristics complicate accurate state capture and restoration, requiring rugged preemption mechanisms to efficiently manage context switching and task resumption.

\vspace{-0.5em}\subsubsection{System-Level Performance Benchmarks} This category includes Dhrystone and CoreMark, two widely adopted benchmarks used to evaluate general-purpose CPU throughput, instruction mix efficiency, and overall control-flow performance in embedded systems~\cite{RISC_dhrystone,RISC_coremark}.
Dhrystone is a synthetic integer benchmark for general-purpose CPU evaluation. It combines arithmetic operations, string processing, and control structures to approximate average embedded workloads~\cite{gal2012exploring}. By contrast, CoreMark is an industry-standard benchmark developed by the embedded microprocessor benchmark consortium (EEMBC) to replace Dhrystone~\cite{gal2012exploring}. It incorporates list processing, matrix computation, finite-state machine operation, and cyclic redundancy check (CRC) validation to characterize CPU efficiency comprehensively. Representing typical CPU-intensive workloads, these benchmarks comprehensively stress architectural state capture, critical for evaluating general-purpose CPU preemption scenarios effectively, thus providing holistic validation for processor-centric preemption mechanisms.

Benchmarks from MachSuite~\cite{benchmark} were adapted to run on a RISC-V soft-core processor and further modified to support full preemption, enabling context switching at any clock cycle. By integrating both PL-based and RISC-V processor-based, our suite enables comprehensive assessment of preemption mechanisms for both dedicated logic and soft-core CPU workloads, ensuring validation across a broad range of real-world scenarios.

\vspace{-1.5em}
\section{Design Configuration of the Proposed Benchmark Suite}\label{Sec:Design_Details}
\textbf{Hardware Setup.} We implemented and tested preemption support within our benchmark suite using the Xilinx Zynq-$7000$ SoC (XC$7$Z$020$) and Vivado $2018.2$ toolchain. The Xilinx Zynq-$7000$ SoC was deliberately selected due to its broad academic adoption, established partial reconfiguration support, and integrated ARM-based processing system (PS) facilitating runtime coordination~\cite{Config_guide}. 

This platform is widely recognized within FPGA research for reliably evaluating multi-tenant scheduling, fast reconfiguration techniques, and hardware resource management, thereby ensuring our results are representative and reproducible across multiple research domains~\cite{vr-zycap, ICAP_Accelerate_3, malik2020isolation, malik2025epoch, karabulut2024themis}. It is also representative of mid-range FPGA platforms~\cite{PR_book}. Moroever, despite its limited resources compared to high-end devices, the XC$7$Z$020$  supports the same partial reconfiguration infrastructure, making it a practical and well-established choice for evaluating reconfigurable systems under preemption. Likewise, using the widely accessible Vivado Design Suite—approximately $51\%$ of the global FPGA market~\cite{l3sbc2024top3}—further enhances reproducibility by providing standardized synthesis, implementation, and testing environments that researchers can readily replicate without specialized resources.


\textbf{Software Setup.} The logic to initiate and manage preemption is implemented on the Zynq-PS, following an approach similar to that proposed in prior work~\cite{malik2025epoch}. The control software is written in C using the Xilinx SDK $2018.2$ and leverages the processor configuration access port (PCAP) to interact with the programmable logic (PL). To safely pause an accelerator, the design utilizes the PS-to-PL clock interface, enabling clean suspension of activity before state capture. The detailed procedures required for safe state saving, including necessary synchronization and precautionary steps, follow the methods described in a prior work and are beyond the scope of this work~\cite{malik2025epoch}. Interested readers are referred to~\cite{malik2025epoch} for further technical details.

\begin{table}[b]
\setlength{\tabcolsep}{7pt}
\centering
\small
\vspace{-2em}
\caption{Total available FPGA resources (LUTs, FFs, BRAMs, DSP units) for designs with one, two, and three slot partitions.}
\vspace{-1em}
\begin{tabular}{|c|c|c|c|c|c|}

\hline
\textbf{\# of slots} & \textbf{Slot\#} & \textbf{LUTs} & \textbf{FFs} & \textbf{BRAMs} & \textbf{DSPs} \\ \hline
$1$                  & $1$             & $11200$       & $22400$      & $80$           & $60$          \\ \hline
\multirow{2}{*}{$2$} & $1$             & $11200$       & $22400$       & $80$            & $60$           \\ \cline{2-6} 
                     & $2$             & $200$       & $400$       & $0$            & $0$           \\ \hline
\multirow{3}{*}{$3$} & $1$             & $800$       & $1600$       & $10$            & $10$           \\ \cline{2-6} 
                     & $2$             & $1600$       & $3200$       & $20$            & $20$           \\ \cline{2-6} 
                     & $3$             & $10000$       & $20000$       & $80$            & $60$           \\ \hline
\end{tabular}
\label{tab:Area_Utilization}
\vspace{-2.0em}
\end{table}
\vspace{-.5em}
\begin{figure}[t!]
    \centering
     \includegraphics[width =\columnwidth]{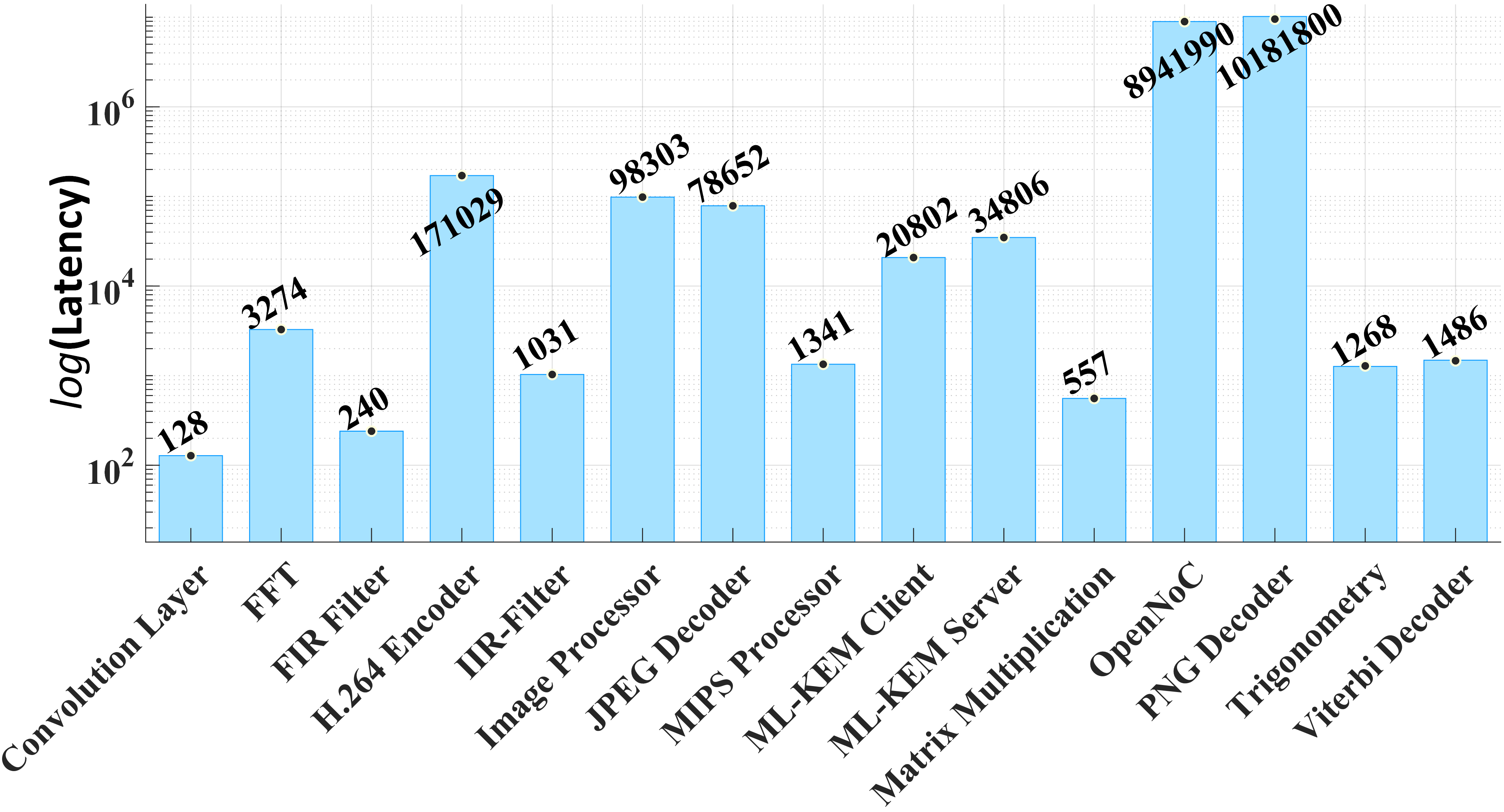}
     \vspace{-2.20em}
        \caption {The figure displays the latency of $15$ PL-based hardware accelerators with preemption support (which we are releasing as open-source). The X-axis shows the benchmark names, while the Y-axis uses a logarithmic scale to present their corresponding latency. We place the (average) latency values—in clock cycles—above each bar to aid interpretation.}
        \label{fig:Latency_PL}
\vspace{-1.50em}
\end{figure}
\vspace{-0.25em}
\subsection{PL-based Hardware Accelerators}
\vspace{-0.25em}
Fig.~\ref{fig:Latency_PL} shows the (average) latency of $15$ PL-based hardware accelerators for which we enabled preemption support\footnote{The suite will be made public upon acceptance of this paper. (Refer \S~\ref{Code})\label{Opensource}}. The X-axis lists the benchmark names, while the Y-axis presents the corresponding latency on a logarithmic scale. We also annotate the average latency values, measured in clock cycles, above each bar to aid interpretation. Table~\ref{tab:PL-based hardware accelerators} \label{PL-based-Hardware-Accelerator}complements this data, summarizing the FPGA resource utilization—lookup tables (LUTs), flip-flops (FFs), block RAMs (BRAMs), and DSP units—for each PL-based hardware accelerator. The evaluation considers configurations where the PL fabric is partitioned into one, two, and three slots. These partitions differ only in allocated area, which can enable a preemption-capable scheduling algorithm to improve resource efficiency by assigning the smallest available slot.

Table~\ref{tab:Area_Utilization} details the slot configurations used in these experiments. This multi-partition setup supports systematic design space exploration and provides an empirical basis for scalability assessment. All benchmarks operate at frequencies between $50$ and $100$ MHz, driven by the PS-side clock, thereby avoiding additional area overhead in line with the methodology of~\cite{malik2025epoch}.
\begin{table}[t!]
\setlength{\tabcolsep}{5.5pt}
\scriptsize
\vspace{-2.0em}
\caption{Resource utilization for PL-based hardware accelerators in terms of LUTs, FFs, BRAMs, and DSP units.}
\vspace{-1.0em}
\label{tab:PL-based hardware accelerators}
\begin{tabular}{|c|c|c|c|c|c|}
\hline
\textbf{Classification}                                                              & \textbf{\begin{tabular}[c]{@{}c@{}}Benchmark\\ Name\end{tabular}}           & \textbf{LUTs} & \textbf{FFs}  & \textbf{BRAMs} & \textbf{DSPs} \\ \hline
\multirow{3}{*}{\begin{tabular}[c]{@{}c@{}}Video\\ Processing\end{tabular}}          & PNG Decoder                                                                 & 2737          & 597           & \textbf{22}    & 0             \\ \cline{2-6} 
& JPEG Decoder                                                                & 2334          & 469           & 5              & 0             \\ \cline{2-6} 
& H.264 Encoder                                                               & 3034          & 1095          & 2              & 2             \\ \hline
\multirow{2}{*}{\begin{tabular}[c]{@{}c@{}}Communication \\ System\end{tabular}}     & Viterbi Decoder                                                             & 39            & 11            & 0              & 0             \\ \cline{2-6} 
& OpenNoC                                                                     & 4980          & 4017          & 0              & 0             \\ \hline
\multicolumn{1}{|r|}{\multirow{2}{*}{Cryptography}}                                  & ML-KEM Server                                                               & \textbf{7016} & 2985          & 3              & 2             \\ \cline{2-6} 
\multicolumn{1}{|r|}{}                                                               & ML-KEM Client                                                               & 7283          & 3002          & 3              & 2             \\ \hline

\multirow{3}{*}{\begin{tabular}[c]{@{}c@{}}Signal\\ Processing\end{tabular}}         & FFT                                                                         & 2508          & \textbf{6096} & 3              & \textbf{32}   \\ \cline{2-6} 
& IIR Filter                                                                  & 74            & 36            & 0              & 0             \\ \cline{2-6} 
& FIR Filter                                                                  & 88            & 41            & 0              & 0             \\ \hline
\multirow{2}{*}{\begin{tabular}[c]{@{}c@{}}Computational \\ Benchmarks\end{tabular}} & MIPS Processor                                                              & 916           & 197           & 8              & 0             \\ \cline{2-6} 
& Trigonometry                                                                & 1242          & 435           & 0              & 0             \\ \hline
\multirow{5}{*}{\begin{tabular}[c]{@{}c@{}}Machine\\ Learning\end{tabular}}          & Image Processor                                                            & 84            & 8             & 0              & 0             \\ \cline{2-6} 
                         & \begin{tabular}[c]{@{}c@{}}Neural Network \\ Convolution Layer\end{tabular} & 2082          & 730           & 0              & 0             \\ \cline{2-6} 
                         & \begin{tabular}[c]{@{}c@{}}Matrix\\ Multiplication\end{tabular}             & 557           & 141           & 108            & 2             \\ \hline

\end{tabular}
\vspace{-2.0em}
\end{table}
\begin{figure}[b!]
\vspace{-1.25em}
    \centering
     \includegraphics[width =\columnwidth]{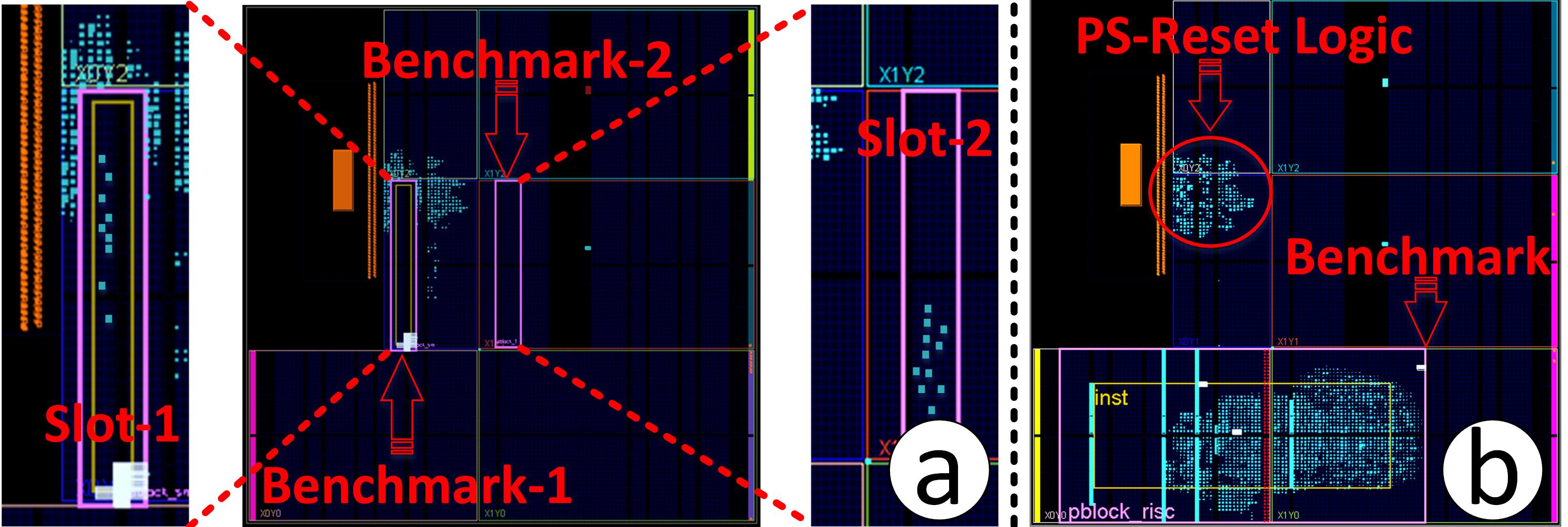}
     \vspace{-1.5em}
        \caption{FPGA floorplan visualization of two deployment configurations: \Circle{a} a two-slot design featuring PL-based hardware accelerators placed over reconfigurable regions spanning clock regions \texttt{X0Y1} and \texttt{X1Y1}; and \Circle{b} a single-slot design hosting a RISC-V–based hardware accelerator distributed across clock regions \texttt{X0Y0} and \texttt{X1Y0}.}
        \label{fig:FPGA-Layout}
\end{figure}
Fig.~\ref{fig:FPGA-Layout}\Circle{a} shows a two-slot FPGA layout where two PL-based hardware accelerators are instantiated on the reconfigurable fabric spanning clock regions \texttt{X0Y1} and \texttt{X1Y1}. In this configuration, both accelerators are clocked using the PS-side clock to avoid area overhead.
\vspace{-0.5em}
\subsection{RISC-V Processor-based Hardware Accelerators}
\begin{figure}[t!]
\vspace{-1.45em}
    \centering
     \includegraphics[width =\columnwidth]{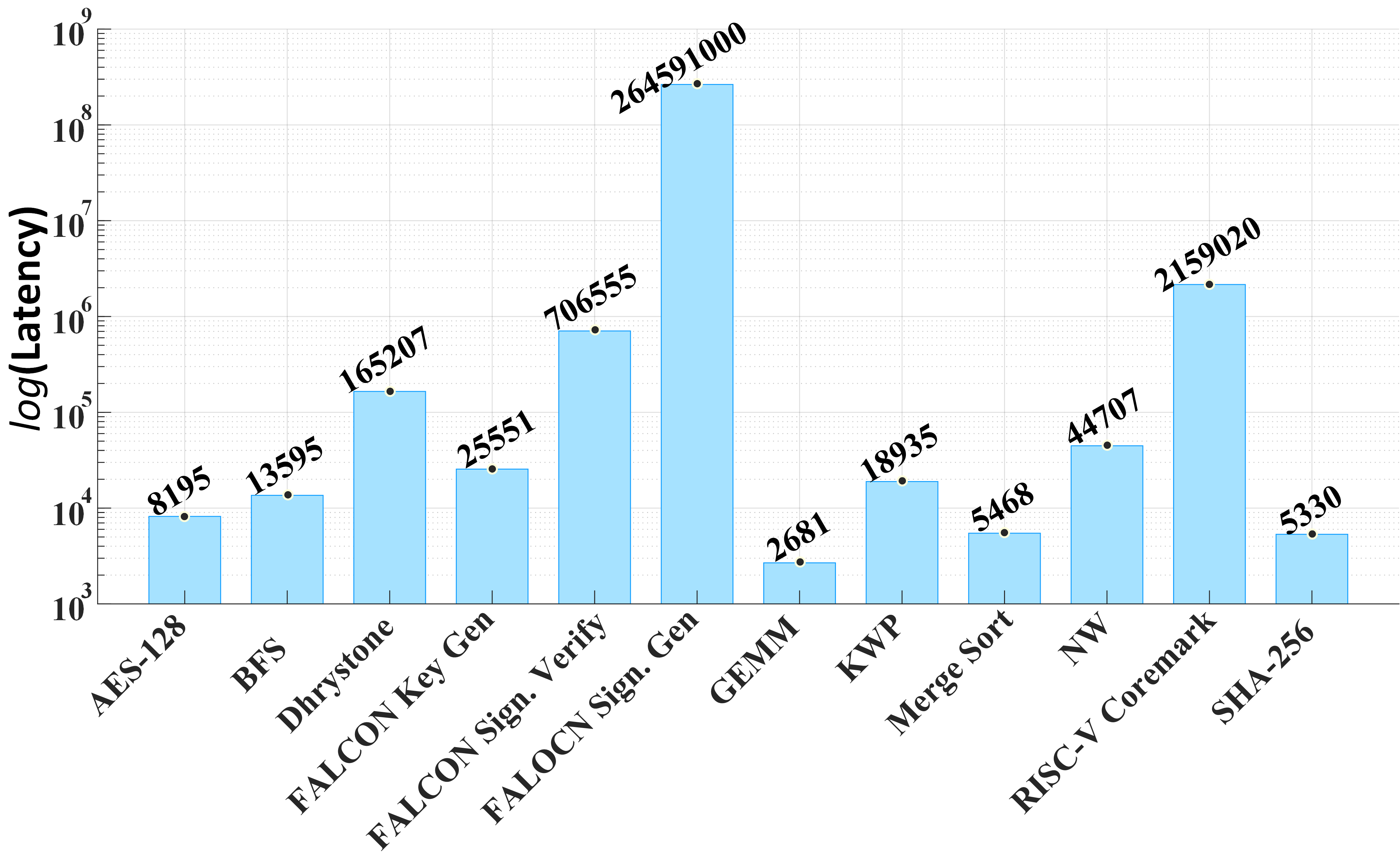}
     \vspace{-2.0em}
        \caption{The latency of $12$ RISC-V processor-based hardware accelerators with preemption support (which we are making open-source). The X-axis shows the benchmark names, while the Y-axis uses a logarithmic scale to present their corresponding latency. We place the (average) latency values—in clock cycles—above each bar to aid interpretation.
        }
        \label{fig:Latency_RISC}
\vspace{-2.0em}
\end{figure}

Fig.~\ref{fig:Latency_RISC} presents the (average) latency of $12$ RISC-V processor-based hardware accelerators that we modified to support preemption\textsuperscript{\ref{Opensource}}. The benchmark names are displayed along the X-axis, while the Y-axis uses a logarithmic scale to represent their execution latencies. To enhance clarity, average latency values in clock cycles are also annotated above each corresponding bar. Table~\ref{tab:RISC-V processor-based hardware accelerators} presents the FPGA resource utilization—LUTs, FFs, BRAMs, and DSP units—for each RISC-V processor-based hardware accelerator. Table~\ref{tab:RISC-V processor-based hardware accelerators} also contains the code size for these accelerators in kilobytes (KB), rounded up to the closest value. Similar to PL-based accelerators, these benchmarks have also been tested by partitioning the PL fabric into one, two, and three slots, respectively. Fig.~\ref{fig:FPGA-Layout}\Circle{b} shows an example of a single-slot design that hosts a RISC-V–based hardware accelerator occupying clock regions, \texttt{X0Y0} and \texttt{X1Y0}. These layouts demonstrate how preemptable accelerator regions can be flexibly mapped across multiple clock regions, allowing for modular and scalable deployment.
\\\indent
Since RISC-V processor-based benchmarks execute entirely in software on the RISC-V core, they exhibit comparable FPGA resource footprints, regardless of the specific benchmark logic. Minor variations in LUT and FF usage arise from differences in compiled code and synthesis tool optimizations. The execution latencies of these benchmarks range from a few thousand clock cycles (\textit{e.g.,} for simple kernels such as \texttt{Sort}) to over two million cycles (\textit{e.g.,} \texttt{CoreMark}), demonstrating a wide span of workload complexities, as shown in Fig.~\ref{fig:Latency_RISC}. This diversity enables this work to evaluate preemption performance across both lightweight and computationally intensive tasks.
\vspace{-0.75em}
\subsection{Verifying Preemption}
To verify functional correctness after state preemption, we implemented a monitoring and validation framework for all accelerators in the benchmark suite. Each accelerator’s output was routed via AXI to the PS for continuous inspection. At randomized intervals, the PS initiated preemption: (i) the accelerator’s state was first saved to DRAM, (ii) the PL slot was then reset to eliminate intermediate data, and finally, (iii) the design was restored from the saved state. The restored accelerator resumed execution from the exact point of interruption, and its outputs were compared against expected results to detect any deviation.

As an example, consider the AES accelerator operating on a partially reconfigurable slot in the PL fabric of a Zynq SoC. The experiment performed one million chained AES encryptions, where the output of encryption `i' served as the input to encryption `i+1'. The PS monitored outputs through the AXI interface and, at randomized intervals, preempted the accelerator by saving its complete state to DRAM. After state saving, the PS reinitialized the PL slot to remove transient data and restored the AES design from DRAM. Execution then resumed until all encryptions were completed. For simplicity, the AES key was fixed to all zeros. The final ciphertext matched a precomputed reference, and identical results across multiple trials confirmed that the state was restored correctly and execution resumed without error.
\begin{table}[t!]
\setlength{\tabcolsep}{3.8pt}
\scriptsize
\vspace{-2.0em}
\caption{Resource utilization for RISC-V processor-based accelerators in terms of LUTs, FFs, BRAMs, DSP units, and the code size in kilobytes (KB), rounded up to the nearest value. }
\vspace{-1.25em}
\label{tab:RISC-V processor-based hardware accelerators}
\begin{tabular}{|c|c|c|c|c|c|c|}
\hline
\textbf{Classification}                                                           & \textbf{\begin{tabular}[c]{@{}c@{}}Benchmark\end{tabular}}  & \textbf{LUTs} & \textbf{FFs}  & \textbf{BRAMs} & \textbf{DSPs} & \textbf{\begin{tabular}[c]{@{}c@{}}Code Size\\(in KB)\end{tabular}} \\ \hline

\multirow{8}{*}{Cryptography}                                                     & AES-128                                                             & 3978          & 2353          & 32             & 3  & 81            \\ \cline{2-7} 
& SHA-256                                                             & \textbf{4070} & 2323          & 32             & 3   & 77            \\ \cline{2-7} 
& \begin{tabular}[c]{@{}c@{}}FALCON \\Key Generate\end{tabular} & 4070          & \textbf{2356} & 32             & 3    & 38           \\ \cline{2-7} 
& \begin{tabular}[c]{@{}c@{}}FALCON \\Sign. Generate\end{tabular}    & 4070          & 2356          & 32             & 3 & 63              \\ \cline{2-7} 
& \begin{tabular}[c]{@{}c@{}}FALCON \\Sign. Verify\end{tabular}      & 4070          & 2356          & 32             & 3 & 35              \\ \hline
\begin{tabular}[c]{@{}c@{}}Machine\\ Learning\end{tabular}                        & GEMM                                                                & 4056          & 696           & 32             & 3 & \textbf{135}              \\ \hline
\multirow{2}{*}{\begin{tabular}[c]{@{}c@{}}Database\\ Operations\end{tabular}}    & BFS                                                                 & 4059          & 700           & 32             & 3   & 103            \\ \cline{2-7} 
& Sort                                                                & 4053          & 680           & 32             & 3    & 115           \\ \hline
\multirow{2}{*}{\begin{tabular}[c]{@{}c@{}}Pattern\\ Recognition\end{tabular}}    & NW                                                                  & 4057          & 685           & 32             & 3   & 116            \\ \cline{2-7} 
& KMP                                                                 & 4053          & 695           & 32             & 3     & 8          \\ \hline
\multirow{3}{*}{\begin{tabular}[c]{@{}c@{}}Performance\\ Benchmarks\end{tabular}} & Dhrystone                                                          & 4055          & 686           & 32             & 3  & 73           \\ \cline{2-7} 
                                                  & \begin{tabular}[c]{@{}c@{}}RISC-V \\ Coremark\end{tabular}         & 4052          & 694           & 32             & 3             & 93                                                             \\ \hline

\end{tabular}
\vspace{-3.0em}
\end{table}
\vspace{-0.5em}
\subsection{Case Study}\label{Case Study}
We now present a case study demonstrating how the proposed benchmark suite can be used to evaluate scheduling policies under realistic preemption overheads. We note that our goal is not to identify the optimal scheduling but to showcase how a scheduling scheme could be mounted on top of our benchmark.
\vspace{-0.5em}
\subsubsection{Round-Robin Scheduling}\label{Round-Robin Scheduling}
To showcase the practical value of our preemption-enabled benchmarks, we evaluate PL allocation using the round-robin (RR) scheduling algorithm using our measured preemption latencies of $0.0311ms$ for context save and $0.0337ms$ for context restore, yielding a total cost of $0.0648ms$ per context switch (see Sec.~\ref {Overhead}), at a clock frequency of $100$MHz. Each accelerator is assumed to occupy a single PL configuration frame, and workloads are sustained so that the PL slot is evicted at every quantum boundary. The objective of this case study is to assess the effect of preemption when scheduling a set of $100$ jobs in a multi-tenant FPGA environment. 
\\\indent
To illustrate this, we analyze two representative scenarios: (i) a batch of $100$ executions of the ML-KEM Server accelerator with a quantum shorter than its execution time, and (ii) a mixed workload in which ML-KEM and an H.264 decoder time-multiplex a single PL slot with a quantum shorter than the H.264 execution time but longer than ML-KEM’s. In both cases, the absence of preemption causes any job exceeding the quantum to lose all progress upon eviction, whereas preemption enables bounded and predictable completion.

In scenario (i), the ML-KEM Server accelerator with a latency of $0.34806ms$ (see Fig.~\ref{fig:Latency_PL}) is scheduled with a quantum of $0.25ms$. Without preemption, the workload cannot complete any execution because the quantum is shorter than the workload's execution time, resulting in unbounded turnaround time. With preemption enabled, each execution spans two quanta and incurs a single context-switch cost, producing an execution time of $2\times0.34806ms + 0.0648ms = 0.41286ms$. The $100$-execution batch, therefore, completes in $100 \times 0.41286$ = $41.286ms$, converting what would otherwise be an infinite turnaround time into a finite, predictable time with minimal per-execution overhead.

In scenario (ii), the quantum is increased to $0.50ms$, and the PL slot is now time-multiplexed between ML-KEM Server ($0.34806ms$) and an H.264 decoder ($1.71029ms$, from Fig.~\ref{fig:Latency_PL}). ML-KEM Server can complete within a single quantum, but H.264 cannot complete without preemption, yielding unbounded turnaround time. With preemption enabled, H.264 completes in four slices—three full 0.50 ms slices and a final 0.21029 ms tail—incurring three context switches for a total of $3 \times 0.0648ms =0.1944ms$ of overhead. Between its slices, RR can schedule one complete ML-KEM Server execution ($0.34806ms$), adding up to $3 \times 0.34806=1.04418ms$ of gap time. The resulting H.264 turnaround time is $1.71029+0.1944+1.04418=2.94887ms$, while ML-KEM Server remains unaffected at $0.34806ms$ per execution.

Across both scenarios, preemption transforms unbounded turnaround time into bounded and predictable completion times. Thus, combined with the fairness guarantees of round-robin scheduling, preemption ensured forward progress in multi-tenant FPGA systems.

\vspace{-0.5em}\subsubsection{Characterizing Preemption Overheads}\label{Characterizing Preemption Overheads}
We now examine the overhead of enabling preemption for the above two representative benchmarks. ML-KEM Server requires $34,806$ cycles per execution, corresponding to $0.34806ms$ at $100$ MHz. By contrast, H.264 has a much larger state footprint (refer Table~\ref{tab:PL-based hardware accelerators} for area) and a latency of $171,029$ cycles ($1.71029ms$). For ML-KEM Server, this overhead is non-negligible relative to its $0.34806ms$ runtime, but unnecessary if the scheduling quantum is large (\textit{e.g.,} $1ms$), since each execution would finish within a single time slice. For H.264, however, preemption is essential. Under a $0.5ms$ quantum, it requires four time slices, incurring three context switches for a total of $0.194ms$ overhead. This adds $\approx11$\% overhead to the baseline runtime but ensures bounded turnaround, whereas without preemption the workload would never complete.

These results show that preemption overhead scales with design complexity. Mid-latency benchmarks such as ML-KEM Server benefit conditionally depending on the scheduler’s quantum, while long-latency benchmarks such as H.264 fundamentally require preemption to make progress. By capturing both behaviors, the benchmark suite provides a practical basis for assessing when preemption is worthwhile and when execution to completion is more efficient.

\vspace{-0.5em}\section{Getting Started and Extending the Suite}\label{Sec:Extending_Bennchmark_Suite}
We now outline the steps to deploy existing benchmarks and how to extend the suite by integrating new designs.

\vspace{-0.75em}\subsection{Step-by-Step Deployment of Preemption Benchmarks and Scheduling Algorithms}

The typical process for deploying and testing a preemption-enabled benchmark on an FPGA using our suite is described here, particularly for evaluating scheduling algorithms that control preemption and context switching.

\begin{enumerate}
    \item \textbf{Reconfiguration Partition Selection:} An existing reconfigurable partition (RP) from the suite is selected by the developer, or a new one is created. If a new RP is created, resource, timing, and interface requirements must be met, and location details such as slice coordinates must be provided.

    \item \textbf{Clock Selection:} The appropriate clock source for the design is set, and the benchmark configuration is updated accordingly.

    \item \textbf{Design Configuration:} The user transfers the source code or synthesized netlist to the FPGA platform. In our work, this typically involves using one of the benchmark modules, referred to as `RTL-IP' (included in the provided open-source code), which is part of our suite. This module is specifically designed and tested to meet the RP’s constraints on logic resources, power consumption, and timing closure for an average user design. By supplying `RTL-IP' alongside other benchmarks, we standardize and simplify deployment, ensuring compatibility with the target platform and enabling consistent evaluation across diverse workloads. This approach also reduces user effort in verifying design compatibility, as the benchmarks are pre-validated against key hardware constraints.

    \item \textbf{Design Deployment:} The scheduling algorithm is integrated and deployed to manage the timing and occurrence of preemption events during execution. This allows different scheduling strategies to be tested with the benchmark.

    \item \textbf{State Capture:} While the design is running, preemption events can be triggered at any time using the suite’s serial-port menu. The `save` command is used to store the current state off-chip.

    \item \textbf{State Restoration:} To resume operation, the `restore` command is invoked via the menu to reload the saved state and continue execution.

\end{enumerate}

\vspace{-0.75em}\subsection{Benchmark Suite Extension}

To foster community-driven growth and continued relevance, our benchmark suite is designed to be extensible. Researchers and practitioners wishing to contribute new benchmarks or adapt existing designs for preemption support can follow these steps:

\begin{enumerate}
    \item \textbf{Benchmark Selection or Development:} Identify or develop a hardware accelerator or software kernel representing a new or emerging application domain.
    \item \textbf{State Element Analysis:} (Optional) Identify all internal states (registers, buffers, memory elements) required for correct pause and resume behavior. This step is primarily for debugging or verifying correct operation pre- and post-preemption.

    \item \textbf{Integration of Preemption Hooks:} Instrument the design with logic to capture and restore its internal state. To this end, the user needs to convey three things to the code running on the PS side:
    \begin{itemize}
        \item The size of the reconfigurable partition (RP) that will host the new accelerator design. \label{Step1}
        \item The clock source for the new accelerator\footnote{This can be either a PL-side clock or a PS-side clock. For simplicity, however, we recommend using the PS-side clock.}.
        \item Assign a non-overlapping memory region in off-chip memory for saving and restoring the state of the new accelerator. The size of this memory region must be chosen according to the size of the RP hosting the accelerator.
    \end{itemize}

    \item \textbf{Validation and Testing:} Develop a testbed to verify that the design produces identical outputs whether it executes uninterrupted or is paused and resumed using preemption.
    \item \textbf{Documentation and Submission:} Document the benchmark’s function, resource requirements, and preemption-specific details. Submit the new benchmark for inclusion via the suite’s open-source repository, following the contribution guidelines.
\end{enumerate}
Through this process, users can deploy existing preemption-enabled designs or contribute new benchmarks to the suite.

\vspace{-1.0em}\section{Discussions}\label{Sec:Discussions}\vspace{-0.25em}
We now summarize the key advantages and contributions of our proposed preemption-enabled benchmark suite.

\vspace{-0.75em}\subsection{Secure Logic and Memory Isolation}\label{Tenant-Isolation}\vspace{-0.25em}
Our benchmark suite uses partial reconfiguration (PR) to partition the FPGA fabric into isolated slots, each dedicated to a specific accelerator. This ensures logic isolation and prevents interference between accelerators. PR allows dynamic preservation, preemption, and restoration of regions without affecting the entire system. Memory isolation is enforced by assigning distinct memory regions in off-chip DRAM to each partition.

To enhance security, our suite employs authenticated encryption—AES-$128$ for encryption and SHA-$256$ for integrity verification—to protect the state and context of each accelerator during state-saving and restoration operations. These measures ensure the confidentiality and integrity of data in multi-accelerator or cloud FPGA environments.
Overall, these mechanisms provide lightweight logic and memory isolation, supporting secure multi-accelerator operations in FPGA-based cloud computing.
Our goal is not to eliminate all possible attack vectors which emerge and grow continuously, but to provide a basic support for standard practice in secure logic and memory isolation.  A full FPGA protection framework could follow the prior literature to extend our effort to be more efficient and/or more comprehensive~\cite{malik2024enabling, drimer2008volatile}.
\vspace{-1.5em}\subsection{Scaling Beyond Fixed Slot Configurations}\vspace{-0.25em}

The proposed benchmark suite is architected for scalability and is not constrained to a fixed number of FPGA slots\footnote{The three-slot configuration shown in Section~\ref{PL-based-Hardware-Accelerator} illustrates an example. The proposed work's architecture is not limited to this configuration and is extensible to accommodate slots as required.}. Unlike prior works that limit preemptable regions to a small, fixed count, this work supports an arbitrary number of slots by managing accelerator state dynamically across multiple contexts~\cite{StateMover,StateMover2}. Each slot is assigned a unique, non-overlapping region in DRAM, with allocation sizes proportionally determined by the resource footprint of the corresponding accelerator, such as the number of LUTs and flip-flops. This design enables consistent, low-overhead context capture and restoration, regardless of system scale, and ensures that preemption remains effective even as the number of active accelerator regions increases.

\vspace{-1.25em}\subsection{Establishing Preemption-Centric Benchmarking}\vspace{-0.25em}

While prior suites, such as OpenDwarfs~\cite{krommydas2016opendwarfs} and MachSuite~\cite{benchmark}, focus on general compute or architectural performance, \emph{this work} explicitly targets enabling preemption for a variety of FPGA workloads. It systematizes a dimension long overlooked—context-aware, interruptible execution across diverse applications. By consolidating a collection of benchmarks and making them preemption-capable, the proposed work moves preemption research beyond isolated demonstrations. Each benchmark supports complete state capture and restoration, enabling systematic evaluation of preemption overheads, correctness, and performance bottlenecks. This shift transforms preemption from an ad hoc engineering effort into a measurable, reproducible research domain, on which now different scheduling algorithms can fairly, consistently, and reproducibly compete.

\vspace{-1.25em}\subsection{Comprehensive FPGA Preemption Evaluation}\vspace{-0.25em}
Motivated by the limitations of prior work, we curated a comprehensive set of $27$ hardware benchmarks to promote broader coverage and generalizability in FPGA preemption research. Unlike earlier studies that often relied on a limited or narrowly focused set of designs, our benchmark suite spans a wide range of application domains and computational characteristics, allowing for a more realistic and thorough evaluation of preemption mechanisms. This diverse collection enables researchers to assess preemption strategies across heterogeneous workloads, thus addressing the gap in standardized and representative benchmarking.

Our benchmark suite includes:
\vspace{-.25em}
\begin{itemize}
\vspace{-.25em}\item Algorithmically intensive kernels (\textit{e.g.,} GEMM, matrix multiplication, H.264 video encoding)
\vspace{-.25em}\item Security-critical accelerators (\textit{e.g.,} AES, ML–KEM)
\vspace{-.25em}\item Memory-irregular workloads (\textit{e.g.}, BFS, KMP)
\vspace{-.25em}\item Streaming and pipelined applications (\textit{e.g.,} FFT, PNG/JPG decoder)
\end{itemize}
\vspace{-.5em}
This suite enables stress-testing of preemption mechanisms under representative real-world workloads, advancing the scope of preemption from mechanism to methodology.

\vspace{-0.5em}\subsection{Ensuring Architectural Generality Beyond Platform Constraints}

This work leverages the Zynq-$7000$ SoC as the demonstration platform, selected deliberately for its balanced support of partial reconfiguration capabilities, integrated on-chip control, and widespread adoption within the academic research community.~\cite{vr-zycap, ICAP_Accelerate_3, malik2020isolation, malik2025epoch, karabulut2024themis,PR_book}. Moreover, the architectural strategies used—clock gating for safe halting, state serialization via PCAP, and memory-mapped interfacing—remain portable to other devices with similar configuration interfaces, such as UltraScale+, Intel Agilex). Re-implementing on larger platforms would increase engineering effort without contributing new methodological insights. Instead, our work establishes a vendor-agnostic foundation that others can extend to different platforms or runtime stacks.

\vspace{-.75em}\subsection{Advancing Shared Research Infrastructure}\vspace{-0.250em}

This work aims to fill a longstanding void in the FPGA community: the lack of a \emph{unified} and extensible benchmarking suite with preemption capabilities. Much like SPEC for CPUs~\cite{henning2002spec} or MLPerf for AI systems~\cite{reddi2020mlperf}, this work provides a reusable infrastructure that supports comparison, validation, and iteration of emerging preemption techniques. This work enables a new class of research inquiries, including:

\begin{itemize}
\vspace{-.25em}\item Preemption-aware scheduling policies for shared FPGAs.
\vspace{-.25em}\item Security analysis of saved state in multi-tenant environments.
\vspace{-.25em}\item CPU-FPGA task migration with context preservation.
\end{itemize}
\vspace{-0.5em}
Thus, this work represents a meaningful advancement in reconfigurable computing, providing a structured foundation for future research on preemption.
\vspace{-1.0em}\subsection{Device Migration}
The benchmark suite developed in this work is portable across various FPGA platforms, including transitions from $7$-Series to UltraScale devices commonly used in multi-tenant cloud environments. Adapting it to UltraScale FPGAs requires minimal modifications, mainly adjusting for the increased number of \texttt{words} per configuration frame (from $101$ to $123$)~\cite{malik2025epoch,ICAPcontext,tapp2015configuration}. Our suite advances FPGA preemption evaluation with several key contributions: it supports scalable configurations beyond fixed FPGA slots, establishes a preemption-focused benchmarking domain, enables evaluation across diverse workloads, promotes shared research infrastructure for FPGA multitasking, and simplifies device migration to newer FPGA families. These enhancements provide a structured, extensible platform that enhances reproducibility and consistency in FPGA preemption research.



\vspace{-1.0em}\subsection{Timing Overhead}\label{Overhead}\vspace{-0.25em}
The FPGA configuration memory is organized into discrete units called \textit{frames}, the smallest addressable portions of the device’s configuration data~\cite{Config_guide}. Each frame contains a fixed number of configuration bits defining the behavior of a specific region within the PL, including logic elements, routing resources, and embedded components. For instance, in Xilinx $7$ Series FPGAs, each frame consists of $101$ $32$-bit words ($101$ $\times$ $32$ = $3232$ bits). The $50^{th}$ word holds the frame CRC bits, and the remaining words store the configuration/settings for LUTs, FFs, BRAMs, DSPs, and other resources\footnote{In a $7$-series FPGA, each LUT is a $6$-input LUT. A \emph{LUT-frame} holds the configuration bits for $50$ LUTs ($3200/50)$ in total, whereas a single \emph{FF-frame} accommodates the configuration of $800$ FF data bits.}~\cite{malik2020isolation,stoddard2016hybrid}.

On Zynq XC$7$Z$020$ SoC using the PCAP interface at $100$ MHz, context save and restore took approximately $ 0.0311$ and $ 0.0337$ ms per frame, respectively~\cite{ZynQ_TRM}. Migrating to newer FPGA families like Xilinx UltraScale may alter these timings due to larger frame sizes—UltraScale frames contain $123$ words—potentially increasing per-frame preemption time proportionally~\cite{ICAPcontext}. Therefore, such variations should be considered when adapting different reconfiguration strategies or deploying our benchmarks across different FPGAs.

Applications requiring frequent partial reconfiguration or rapid context switching can benefit from reducing per-frame save and restore time. By quantifying this overhead, we provide realistic, generalizable insights for diverse workloads, supporting informed FPGA design decisions. While timing overhead is important, optimizing save and restore latencies is beyond the scope of this work. Our goal is to deliver a diverse benchmark suite with preemption capabilities; reducing preemption overhead remains future work.

\vspace{-1.0em}\subsection{To Preempt or Not to Preempt}\label{Sec:Preemption Overhead}\vspace{-0.25em}
Sec.~\ref{Case Study} demonstrates that preemption is not a one-size-fits-all solution; its benefit depends on accelerator characteristics and the chosen scheduling policy. Inherently, preemption is most valuable for applications with large state sizes, high latency, and long execution times. Such workloads are likely to benefit further when combined with a scheduling algorithm that allows a high-priority task to interrupt a lower-priority one. Our diverse benchmark suite is designed to encompass this full spectrum. While we acknowledge some benchmarks have diminished benefits from preemption, we include them to enable comprehensive evaluation and pave the way for future work that may reduce context-switch overhead. The effectiveness of preemption hinges on a delicate balance between the scheduling quantum, policy, and preemption timing. Our benchmarks provide a methodical framework for exploring these critical relationships and tailoring scheduling strategies to different hardware and application needs.

\vspace{-1.0em}
\section{Conclusion}\label{Sec:Conclusion}\vspace{-0.5em}
This work presented an open-source benchmark suite designed to evaluate hardware preemption in FPGA-based systems. By integrating a diverse set of preemption-capable benchmarks across multiple domains, our work enables systematic exploration of preemption overheads, design trade-offs, and architectural implications. While demonstrated on the Zynq-$7000$ platform, the techniques employed remain applicable to a broad class of reconfigurable devices. As hardware multitasking and multi-tenancy become increasingly relevant in FPGA deployments, our work offers a practical foundation for evaluating and advancing preemption-aware runtime systems.

\vspace{.5em}
\noindent \textbf{Code review.} \label{Code}The benchmark suite is available online under the \href{https://github.com/aamalik3/Benchmarks/}{MIT license}\footnote{https://github.com/aamalik3/Benchmarks/}.

\small{\fontsize{1}{1}\selectfont{\bibliographystyle{ACM-Reference-Format}}
\bibliography{references.bib}

\end{document}